%%%%%%%%%%%%%%%%%%%%%%%%%%%%%%%%%%%%%%%%%%%%%%%%%%%%%%%%%%%%%%%%%%%%%%%%%%%%
%%%%%  FROM PCM to KDV AND BACK 
%%%%%  
%%%%%  J.M. EVANS 
%%%%%
%%%%%  PLAIN TEX
%%%%%%%%%%%%%%%%%%%%%%%%%%%%%%%%%%%%%%%%%%%%%%%%%%%%%%%%%%%%%%%%%%%%%%%%%%%%
%\magnification=\magstep1
\openup 1\jot
\rightline{PUPT-1956}
\rightline{DAMTP-2000-141}
\rightline{hep-th/0103250}
\vskip 40pt
\centerline{\bf FROM PCM TO KDV AND BACK}
\vskip 20pt

\centerline{Jonathan M. Evans}
\vskip 5pt
\centerline{\it Joseph Henry Laboratories, Princeton University,
Princeton NJ 08544, USA}
\vskip 1pt
\centerline{\it DAMTP, University of Cambridge, Silver Street, Cambridge
CB3 9EW, UK}
\vskip 15pt
\centerline{Contribution to the NATO-ASI meeting} 
\centerline{\it Integrable Hierarchies and Modern Physical Theories} 
\centerline{University of Illinois at Chicago, July 2000} 
\vskip 40pt

\noindent
This is a summary of progress made  
[1-4] in understanding the occurrence and properties 
of local, conserved, commuting charges in non-linear sigma-models, 
including principal chiral models (PCMs) and WZW models.
In each PCM or WZW model with target manifold a compact 
Lie group $G$, there are infinitely many commuting charges 
whose currents have the form 
${\cal K}_m \! =  k_{a_1 a_2 \ldots a_m} j^{a_1} j^{a_2} \! \ldots
j^{a_m} $.
The underlying fields $j^a$ take values in the Lie algebra ${\bf g}$
(indices $a$ refer to some basis)
and each tensor $k_{a_1 a_2 \ldots a_m}$ is totally symmetric 
and $G$-invariant. 
The spins of the corresponding conserved charges $s=m-1$ 
run over the exponents of the algebra modulo its Coxeter number $h$. 
Such patterns of spins are familiar from the context of affine Toda field 
theory [5] and these similarities offer a natural explanation of some 
otherwise mysterious common properties of PCM and affine Toda 
S-matrices [1].

Initial investigations [1] focussed on PCMs 
based on classical groups $G$ and established that  
currents and symmetric tensors with the required properties 
could be defined by a formula
${\cal K}_m = \det ( 1 - \mu j^a t_a )^{s/h} |_{\mu^{s+1}}$,
where the generators $t_a$ belong to the defining representation 
of ${\bf g}$.
Commutation of the resulting charges depends upon some intricate 
algebraic identities satisfied by the $k$-tensors which we shall refer
to as the {\it commutation conditions\/}.
These findings were subsequently extended to incorporate the effects of 
WZ terms and supersymmetry [2] and analogous results were also established 
for sigma-models whose target manifolds are symmetric spaces
$G/H$ [3], but again only for $G$ and $H$ classical groups. 
The general case, including exceptional groups, has been treated 
quite recently [4] by making use of a direct link 
between the $k$-tensors and the Drinfeld-Sokolov modified 
KdV (DS/mKdV) hierarchies [6].  

The first step in establishing this link is to show that
all relevant attributes of the $k$-tensors can be understood 
as properties of the {\it restricted\/} tensors $k_{i_1 i_2 \ldots i_m}$
defined just on the Cartan subalgebra (CSA)
of ${\bf g}$ (indices $i$ refer to the CSA). 
It is well-known that any symmetric invariant tensor 
is completely determined by its restriction and that the remnant of 
$G$ which fixes the CSA is the Weyl group; moreover,
any Weyl-invariant tensor on the CSA can be extended
uniquely to a $G$-invariant tensor on ${\bf g}$. 
It is less obvious that the key commutation conditions of specific 
interest to us can be simply expressed in terms of restricted tensors 
alone, but this is in fact the case [4].

Recall next that the DS/mKdV hierarchy associated with 
a Lie algebra ${\bf g}$ [6] involves currents of the form
${\cal H}_m = h_{i_1 i_2 \ldots i_m} u^{i_1} u^{i_2} \! \ldots u^{i_m}
+ ({\rm derivative~terms})$, where the fields $u^{i}$ live in
the CSA of ${\bf g}$. The explicit term in ${\cal H}_m$ 
contains a certain Weyl-invariant tensor $h_{i_1 i_2 \ldots i_m}$, 
while the additional, implicit terms involve derivatives of the fields.
Drinfeld and Sokolov proved that the charges corresponding to 
these currents always commute. 
It is possible to deduce from this that 
the $h$-tensors must satisfy precisely the same {\it commutation
conditions\/} already mentioned, notwithstanding the  
complications arising from the extra derivative terms.

We can conclude, therefore, that after extension from the CSA 
to all of ${\bf g}$, the $h$-tensors of the DS hierarchies 
provide new definitions of $k$-tensors with exactly 
the properties we require for commuting charges, for any compact 
Lie algebra, whether classical or exceptional.
Careful examination [4] also reveals that whenever concrete 
formulas are available for the DS currents [6], these do indeed 
yield expressions 
for $h$-tensors which are equivalent to those 
obtained from our previous formula for $k$-tensors given above, 
so that the old and new definitions
are entirely consistent.
The new approach can be applied to all
symmetric space sigma-models too [4].

The commuting quantities in the DS/mKdV hierarchies are naturally 
identified with the conserved charges of affine Toda field theories 
(the fields $u^i$ being derivatives of the usual Toda fields).
The relationships sketched here 
underscore the fact that these two prominent  
categories of integrable Lagrangian field theories, namely sigma-models
and Toda theories, have much more in common than might initially be 
supposed. Although the two classes of models appear 
to represent radically different ways of introducing interactions 
amongst collections of fields, we have found that they nevertheless
contain a common core, each exhibiting commuting local charges based on
precisely the same sets of Weyl-invariant tensors.
\vskip 10pt
\noindent
Acknowledgements: My warm thanks to the organizers of the 
conference, Henrik Aratyn and Alexander Sorin, for the invitation to 
speak. Much of the research described here was carried out while
spending a stimulating and very enjoyable year at Princeton University;
I am grateful to all in the Physics Department, but 
especially Erik Verlinde, for making this possible. 
The work was supported by NSF grant PHY98-02484 and 
by a PPARC Advanced Fellowship.
\vskip 20pt

\noindent
\centerline{\bf References}
\vskip 15pt
\noindent
[1] 
J.M.~Evans, M.~Hassan, N.J.~MacKay, and A.J.~Mountain,
%\hfil \break
{\it Local conserved charges in principal chiral models},
%\hfil \break
Nucl.~Phys.~{\bf B561} (1999) 385-412; {\tt hep-th/9902008}.
\vskip 5pt
\noindent
[2] 
J.M.~Evans, M.~Hassan, N.J.~MacKay, and A.J.~Mountain,
%\hfil \break
{\it Conserved charges and supersymmetry in principal chiral and WZW models},
%\hfil \break
Nucl.~Phys.~{\bf B580} (2000) 605-646; {\tt hep-th/0001222}. 
\vskip 5pt
\noindent
[3]
J.M.~Evans and A.J.~Mountain,
%\hfil \break
{\it Commuting charges and symmetric spaces},
%\hfil \break
Phys.~Lett.~{\bf B483} (2000) 290-298; {\tt hep-th/0003264}. 
\vskip 5pt
\noindent
[4]
J.M.~Evans, 
%\hfil \break
{\it Integrable sigma-models and Drinfeld-Sokolov hierarchies\/};
%\hfil \break 
{\tt hep-th/0101231}. 
\vskip 5pt
\noindent
[5]
E.~Corrigan,
%\hfil \break
{\it Recent developments in affine Toda quantum field theory},
%\hfil \break
Lectures at Banff Summer School 1994; 
{\tt hep-th/9412213}.
\vskip 5pt
\noindent
[6]
V.G.~Drinfeld and V.V.~Sokolov, 
%\hfil \break
{\it Lie algebras and equations of Korteweg-deVries type\/}, 
%\hfil \break
J.~Sov.~Math.~{\bf 30} (1985) 1975-2036.

\bye